
\magnification=1200

\baselineskip=12pt

\font\ftit=cmbx10

\parskip=6pt
\parindent=2pc

\font\titulo=cmbx10 scaled\magstep1

\def\qed{{\vrule height 7pt width 7pt depth 0pt}\par\bigskip}

\def\section#1{\vskip 1.5truepc plus 0.1truepc minus 0.1truepc
	\goodbreak \leftline{\titulo#1} \nobreak \vskip 0.1truepc
	\indent}
\def\frc#1#2{\leavevmode\kern.1em
	\raise.5ex\hbox{\the\scriptfont0 $ #1 $}\kern-.1em
	/\kern-.15em\lower.25ex\hbox{\the\scriptfont0 $ #2 $}}




\def\IR{{\rm I\!R}}  
\def\IN{{\rm I\!N}}  

\newbox\pmbbox
 \def\pmb#1{{\setbox\pmbbox=\hbox{$#1$}%
\copy\pmbbox\kern-\wd\pmbbox\kern.3pt\raise.3pt\copy\pmbbox\kern-\wd\pmbbox
\kern.3pt\box\pmbbox}}


\font\cmss=cmss10
\font\cmsss=cmss10 at 7pt

\def\IZ{\relax\ifmmode\mathchoice
{\hbox{\cmss Z\kern-.4em Z}}{\hbox{\cmss Z\kern-.4em Z}}
{\lower.9pt\hbox{\cmsss Z\kern-.4em Z}}
{\lower1.2pt\hbox{\cmsss Z\kern-.4em Z}}\else{\cmss Z\kern-.4em Z}\fi}







\centerline{{\ftit Chaotic Dynamics of Binary Systems}
\footnote{*}{This work is supported in part by CONACyT grant
400349-5-1714E and by the Association G\'en\'erale pour la
Coop\'eration et le D\'eveloppement \break (Belgium).}}

\vskip 0.5pc

\centerline{Henri Waelbroeck and Federico Zertuche${}^{\dagger}$}

\vskip 0.5pc

\centerline{Instituto de Ciencias Nucleares, UNAM} \centerline{Apdo.
Postal 70-543, M\'exico, D.F., 04510 M\'exico.} \centerline{e-mail:
hwael@roxanne.nuclecu.unam.mx}

\vskip 1.0pc

\centerline{${}^{\dagger}$Instituto de Investigaciones en
Matem\'aticas Aplicadas} \centerline{y en Sistemas, UNAM, Secci\'on
Cuernavaca, A.P. 48-3} \centerline{62251 Cuernavaca, Morelos,
M\'exico.} \centerline{e-mail: zertuche@ce.ifisicam.unam.mx}

\baselineskip=22pt

\vskip 1.5pc

\centerline { Abstract}

{\leftskip=1.5pc\rightskip=1.5pc\smallskip\noindent We propose a
theory of chaos for discrete systems, based on their representation
in a space of ``binary histories'', $ {\cal B^{\infty}} $. We show
that $ {\cal B^{\infty}} $ is a metrizable Cantor set which embeds
the attractor $\Lambda$, itself also a Cantor set.
	\smallskip}

\vfill\eject

\vskip 1.5pc

\section{1. Introduction.}

 Most recent results on discrete systems concern states at or near
equilibrium. Martinelli and Oliveri proved that Hamiltonian spin
systems relax exponentially, under the strong mixing
assumption~${}^{[1]}$, and Comets and Neveu showed that fluctuations
of thermodynamic variables in the Sherrington Kirkpatrick model are
Gaussian processes~${}^{[2]}$. The relation of ground states to the
vacuum of quantum field theories is well known~${}^{[3]}$; for
example Cecotti and Vafa proved their equivalence in the case of the
Ising models and supersymmetric field theories in dimension $d =
2$~${}^{[4]}$.

 Less is known about their behaviour far from equilibrium. Ising
models can display interesting dynamical behaviour, as Neves and
Schonmann showed in studies of the transition from a metastable to a
stable equilibrium~${}^{[5]}$ in $d = 2$. Asymmetric Hopfield
networks have transitions to disordered dynamical phases~${}^{[6]}$.
Also, Cellular automata have non-trivial dynamics; soliton solutions
were studied by Bobenko, Bordemann, Gunn and Pinkall~${}^{[7]}$. One
of the more complex dynamical behaviours in Wolfram's
classification~${}^{[8]}$ has been studied empirically as a form of
``chaos''~${}^{[9]}$, but it has never been entirely clear how this
is related to chaotic dynamics in Euclidean space.

 As is well-known, Markovian dynamics on a finite state space cannot
be chaotic since every orbit must eventually fall onto a finite
limit-cycle. However, one would like to have a framework for chaos
which allows one to decide whether the complex dynamics of finite
systems is {\it approximately} chaotic.  Unfortunately, for most
finite systems there is no convenient quasi-representation in terms
of real variables. There are different points of view on this
problem, ranging from the fundamentalist, which concludes that a
finite system cannot be viewed as approximately chaotic, to the
liberal, which reduces the definition of chaos to sensitive
dependence on initial conditions and the exponential growth of the
limit-cycle period with the size of the system.

 Our feeling is that ``chaos'' should not be limited to real
variables, as these are idealizations of a reality which could be
viewed equally well in terms of finite state spaces. Indeed, the fact
that most real variables have infinite algorithmic
information~${}^{[10]}$ is not satisfactory from a physicist's point
of view. Yet some form of idealization is necessary to define
``chaos'' rigorously.

 Our purpose in this article is to propose a different idealization,
inspired from symbolic dynamics~${}^{[11, 12]}$, which is well suited
to finite systems with a natural binary network representation
(including Ising models, spin glasses and cellular automata).

 Instead of specifying the coordinates of a point with infinite
accuracy, we will assume that one is given the $N-bit$ model of the
system at every past tick of a clock, to the infinitely remote past:
the {\it state} of a binary system is given by
$$ S = \{ {\bf S} \left( 0 \right), {\bf S} \left( - 1
	\right) , \cdots , {\bf S} \left( - n \right) , \cdots \}, $$
where $ {\bf S} \left( - n \right) $ is a binary vector with
components $ S^i\left( -n \right) \in \{ 0, 1 \} $; $i = 1, \cdots, N
$. The space of such binary histories will be denoted by $ {\cal
B^{\infty}} $ [Figure 1].

	The approximation which makes this concept practical, akin to
the $ 128-bit $ version of floating-point variables, is the
truncation of the binary history to the $ n $ most recent steps in
the past. This truncation is valid if the difference between states
with the same history for the first $ n $ steps belongs to a small
neighborhood of the origin. We will formalize this demand in the
definition of the ``semicausal topology'', which will be the keystone
of our construction.

 Thus, order $n$ Markow chains on finite spaces can have
quasi-chaotic dynamics if $n$ is large, and be precisely chaotic in
the $n \to \infty$ limit. Cellular automata and asymmetric spin
glasses are $n = 1$ processes, so they do {\it not} provide good
models of chaos. This fact manifests itself in the lack of an
invariant attracting set where the dynamics is topologically
transitive.

 We begin by reviewing definitions for chaotic maps on real phase
spaces~${}^{[12]}$. Let $ f: \IR^d \rightarrow \IR^d $, $ d \in
\IZ^{+} $ be a contiuous map, such that $ x_{n + 1} = f \left( x_n
\right) $.

\noindent {\it Definition 1}: $ f: \IR^d \rightarrow \IR^d $ has {\it
sensitive dependence on initial conditions} on $ {\cal A} \subset
\IR^d $ if $ \exists $ $ \delta > 0 $ $ \ni $ $ \forall $ $ x \in
{\cal A} $ and $ \forall N \left( x \right) $ (neighborhood of $ x $)
$ \exists $ $ y \in N \left( x \right) $ and $ n \in \IN $ $ \ni $ $
\vert f^n \left( x \right) - f^n \left( y \right) \vert > \delta $
{\footnote{ $ \dagger $ }{ $ \IN = \left\{ 0, 1, 2, \dots \right\} $
}}.

\noindent {\it Definition 2}: $ f: \IR^d \rightarrow \IR^d $ is {\it
topologically transitive} on $ {\cal A} \subset \IR^d $ if for any
open sets $ U $, $ V $ $ \subset {\cal A} $ $ \exists $ $ n \in \IZ $
$ \ni $ $ f^n \left( U \right) \cap V \not= \emptyset $.

\noindent {\it Definition 3}: Let $ {\cal A} \subset \IR^d $ be a
compact set. $ f: {\cal A} \rightarrow {\cal A} $ is {\it chaotic} on
$ {\cal A} $ if $ f $ has sensitive dependence on initial conditions
and is topologically transitive on $ {\cal A} $.

\noindent {\it Definition 4}: A closed and connected set $ {\cal M}
\subset \IR^d $ is called a {\it trapping region} if $ f \left( {\cal
M} \right) \subset {\cal M} $.

\noindent {\it Definition 5}: The map $ f $ has a {\it chaotic
attractor} $ \Lambda \subset \IR^d $ if $ \Lambda $ is a compact set
on which $ f $ is chaotic and exists a trapping region $ {\cal M} $
such that $ \Lambda = \bigcap\limits_{n \geq 0} f^n \left( {\cal M}
\right) $.

	Of the five definitions above, only the first uses the
Euclidean metric explicitly. Since the property of ``chaos'' is
topologically invariant, we will use a topological definition of
``sensitivity to initial conditions'':

\noindent {\it Definition 1bis}: $ f: \IR^d \rightarrow \IR^d $ has
{\it sensitive dependence on initial conditions} on $ {\cal M}
\subset \IR^d $ if there exists a field of neighborhoods $ {\cal
N}(x) $, i.e., a function from the trapping region to the continuous
topology on $ \IR^d $, $ {\cal N} : {\cal M} \to \tau({\cal M}) $,
such that
	$$ \forall x \ \exists y \in {\cal N} \left( x \right) \cap
	{\cal M}, \ n \in \IN \ni f^n \left( y \right) \notin {\cal
	N} \left( f^n \left( x \right) \right) $$
The metrical definition (1) is recovered if one requires that the
neighborhoods $ {\cal N} \left( x \right) $ be $ \delta $-balls
centered at $ x $.

\

\section{2. The Space of Binary Histories $ {\cal B^{\infty}} $.}

	Let ${\bf S} = (S_1(t), S_2(t), \cdots, S_N(t))$ be an
$N-bit$ binary model of the state of the system at time $t$. We
denote the space of the $ 2^N $ possible binary states by $ {\cal B}
= \left\{ {\bf S} \right\} $ and the infinite set of binary histories
of the system by
	$$ {\cal B^{\infty}} = \left\{ S = \left( {\bf S} \left( 0
	\right), {\bf S} \left( - 1 \right),... \right) \right\}.
	\eqno(1) $$

	We endow $ {\cal B^{\infty}} $ with a topology such that
near-neighbors in $ {\cal B^{\infty}} $ have similar binary states in
the recent past.

\noindent {\it Definition 6}: A {\it semicausal topology} on $ {\cal
B^{\infty}} $ with index $\Delta \in \IN$ is a topology generated by
a base whose elements $ {\cal N}_{n}^{\Delta} \left( S \right) $, $ S
\in {\cal B^{\infty}} $, $ n \in \IN $ satisfy:

\noindent {\it i)} $ S' \in {\cal N}_{n}^{\Delta} \left( S \right)
\Longrightarrow \forall m < n, {\bf S'} \left( - m \right) = {\bf S}
\left( - m \right) $.

\noindent {\it ii)} $ {\bf S'} \left( - m \right) = {\bf S} \left( -
m \right) \forall m < n + \Delta \Longrightarrow S' \in {\cal
N}_{n}^{\Delta} \left( S \right) $.

	Note that if $ S' \in {\cal N}_{n}^{\Delta} \left( S \right)
$ it may or may not have the same binary states $ {\bf S'} \left( - m
\right) $ of $ S $ in the range $ n \leq m < n + \Delta $. One
semicausal topology differs from another in which differences are
allowed between $ {\bf S'} \left( - m \right) $ and $ {\bf S} \left(
- m \right) $ in this range, for $ {\bf S'} \in {\cal N}_{n}^{\Delta}
\left( S \right) $ [Figure 2].

\noindent It is easy to check that the basis sets $ {\cal
N}_{n}^{\Delta} \left( S \right) $ satisfy the

\noindent{\it Property 1}: $ {\cal N}_{n + \Delta}^{\Delta} \left( S
\right) \subset {\cal N}_{n}^{\Delta} \left( S' \right) $ $ \forall
S' \in {\cal N}_{n + \Delta}^{\Delta} \left( S \right) $.

\vskip 0.5cm

\noindent {\it Definition 7}: The {\it causal} topology is a
semicausal topology with $ \Delta = 0 $. The base elements $ {\cal
N}_n^0 \left( S \right) $ are uniquely defined by {\it i)} and {\it
ii)} above.

	From here on we will assume that $ {\cal B^{\infty}} $ is
equipped with a semicausal topology. To simplify the notation we will
drop the $ \Delta $ in the notation and write the basis elements
simply as $ {\cal N}_{n} \left( S \right) $. The following
proposition is easy to verify:

\noindent {\it Property 2}: $ {\cal B^{\infty}} $ is a boolean
algebra (a ring with an idempotent cross operation) under the logical
operations
	$$ XOR = +, \eqno(2) $$
	$$ AND = \times \eqno(3) $$
performed on each bit in the
infinite binary chain $ S $. The ``zero'' is the element
	$$ 0 = \left( {\bf 0,0,0,0,...} \right), \eqno(4) $$
and the identity is the element
	 $$ e = \left( {\bf 1,1,1,1,...} \right). \eqno(5) $$
The addition operation is nilpotent
	$$ S + S = 0  \eqno(6) $$
and the multiplication is idempotent
	$$ S \times S = S. \eqno(7) $$

\noindent {\it Definition 8}: The sequence $ \left\{ S_k \right\} $,
$ k \in \IN $ is a {\it Cauchy net} if
	$$ \forall n \in \IN \ \exists k_0 \in \IN \ \ni
	S_{k} - S_{k'} \in {\cal N}_n \left( 0 \right) \ \forall k, k' >
	k_0, $$
see Ref.~[13].

\

\noindent {\it Theorem 1}: $ {\cal B^{\infty}} $ is complete.

\noindent {\it Proof}: We need to show that any Cauchy net of
elements in $ {\cal B^{\infty}} $ converges to an element in $ {\cal
B^{\infty}} $. Let $ \left\{ S_k \right\} $ be a Cauchy net.   From
definition 8, $ {\bf S}_{k'} \left( - m \right) = {\bf S}_{k} \left(
- m \right) $ $ \forall m < n $ and $ k, k' > k_0 $.  Let us
construct $ S $ from $m = 0$ down to $ m = n - 1 $ such that $ {\bf
S} \left( - m \right) = {\bf S}_k \left( - m \right) $ $ \forall m <
n $; by incrementing $n$ this construction leads to a unique binary
history $ S $. The sequence $ \left\{ S_k \right\} $ converges to $ S
$ which by construction is of the form (1), so that $ S \in {\cal
B^{\infty}} $. \hfill \qed

\noindent {\it Theorem 2}: $ {\cal B^{\infty}} $ is Hausdorff.

\noindent {\it Proof}: Let $ S \not= S' $, then $ \exists n \in \IN $
$ \ni $ $ {\bf S} \left( - n \right) \not= {\bf S'} \left( - n
\right) $. The neighborhoods $ {\cal N}_{n + 1} \left( S \right) $
and $ {\cal N}_{n + 1} \left( S' \right) $ are disjoint. \hfill \qed

\noindent {\it Theorem 3}: $ {\cal B^{\infty}} $ is perfect.

\noindent {\it Proof}: A set is perfect if it is closed and every
point is an accumulation point. $ {\cal B^{\infty}} $ is closed
because it is the total space so it is both open and closed. Let $ S
\in {\cal B^{\infty}} $ and consider the sequence $ \left\{ S_k
\right\} $ given by
	$$ {\bf S}_k \left( - n \right) = {\bf S} \left( - n \right),
	\forall n < k $$
and
	$$ {\bf S}_k \left( - n \right) = {\bf S} \left( - n \right)
	+ {\bf 1}, \forall n \geq k, $$
where $ {\bf 1} $ is given by (5). This sequence converges to $ S $
and by construction $ S_k \not= S \ \forall k $.  Thus, $ S $ is an
accumulation point. \hfill \qed

\noindent {\it Lemma 1}: If $ S' \notin {\cal N}_n \left( S \right) $
then $ {\cal N}_n \left( S \right) \cap {\cal N}_m \left( S' \right)
= \emptyset $ $ \forall m \geq n + \Delta $.

\noindent {\it Proof}: From definition 6, if $ S' \notin {\cal N}_n
\left( S \right) $ then $ {\bf S'} \left( - k \right) \not= {\bf S}
\left( - k \right) $ for some $ k < n + \Delta $. Let $ S_0 \in {\cal
N}_m \left( S' \right) $ then $ {\bf S}_0 \left( - k \right) = {\bf
S'} \left( - k \right) $ $ \forall k < n + \Delta \leq m $, which
implies that $ S_0 \notin {\cal N}_n \left( S \right) $. \hfill \qed

\noindent {\it Lemma 2}: The complement $ {\cal N}_n^c \left( S
\right) $ of any neighborhood is an open set.

\noindent {\it Proof}: Due to lemma 1, $ {\cal N}_n^c \left( S \right)
$ can be expressed by
	$$ {\cal N}_n^c \left( S \right) = \bigcup\limits_{S' \in
	{\cal N}_n^c \left( S \right)} {\cal N}_{n + \Delta} \left(
	S' \right), $$
which is a union of open sets, then $ {\cal N}_n^c \left( S \right) $
is an open set. \hfill \qed

\noindent {\it Theorem 4}: $ {\cal B^{\infty}} $ is totally
disconnected.

\noindent {\it Proof}: We must show that the connected component $
{\cal C} \left( S \right) $ of each $ S \in {\cal B^{\infty}} $
consists of just the point $ S $~${}^{[14]}$.

	By contradiction: Let $ S' \in {\cal C} \left( S \right) $
with $ S' \not= S $. Then $ S $ and $ S' $ differ in at least one
binary state: $ {\bf S} \left( - n \right) \not= {\bf S'} \left( - n
\right)$. Then $ S' \notin {\cal N}_{n + 1} \left( S \right) $. Now,
	$$ {\cal B^{\infty}} = {\cal N}_{n + 1} \left( S \right)
	\cup {\cal N}_{n + 1}^c \left( S \right) $$
is a separation of $ {\cal B^{\infty}} $ because by lemma 2 bought of
them are disjoint non-empty open sets. This implies that $ {\cal C}
\left( S \right) \subset {\cal N}_{n + 1} \left( S \right) $ but $ S'
\notin {\cal N}_{n + 1} \left( S \right) $ and we have a
contradiction.  \hfill \qed

\noindent {\it Theorem 5}: $ {\cal B^{\infty}} $ is compact.

	Before proving the theorem let us first prove three lemmas.

\noindent {\it Lemma 3}: The number of distinct sets $ {\cal N}_n
\left( S \right) $ $ \forall S \in {\cal B^{\infty}} $ and $ n $
fixed, is finite.

\noindent {\it Proof}: For any $ S \in {\cal B^{\infty}} $,
$ {\cal N}_n \left( S \right) $ is totally defined by
specifying the first $ n + \Delta $ binary states and the list of
which differences are allowed between $ {\bf S'} \left( - m \right)
$ and $ {\bf S} \left( - n' \right) $ in the range $ n \leq m < n +
\Delta $, in all a finite amount of information for any fixed $ n $.
\hfill \qed

\noindent {\it Lemma 4}: Let $ {\cal V} = \left\{ U_i \right\} $ be
an open covering of $ {\cal B^{\infty}} $ such that there exists a $
n_0 \in \IN $ such that $ \forall S \in {\cal B^{\infty}} $ $ \exists
m \leq n_0 $ $ \ni $ $ {\cal N}_m \left( S \right) \subset U_i $ for
some $ U_i \in {\cal V} $. Then there exists a finite subcovering of $
{\cal V} $.

\noindent {\it Proof}: $ \forall S \in {\cal B^{\infty}} $ let $
U_\alpha \in {\cal V} $ be such that $ S \in U_\alpha $ and $ {\cal
N}_m \left( S \right) \subset U_\alpha $ with the minimum possible $
m $. By lemma 3, the number of distinct $ {\cal N}_m \left( S \right)
$ with $ m \leq n_0 $ is finite, so the set of such base elements $
\left\{ {\cal N}_m \left( S \right) \right\} $, is a finite covering.
Since for each $ {\cal N}_m \left( S \right) $ there is an associated
$ U_\alpha $ and $ {\cal N}_m \left( S \right) \subset U_\alpha $,
$\left\{ U_\alpha \right\}$ is a finite subcovering of $ {\cal
B^{\infty}} $. \hfill \qed

\noindent {\it Lemma 5}: If $ g: {\cal N}_n \left( S \right)
\longrightarrow \IN $ is a non-bounded function, then there exists $
{\cal N}_m \left( S' \right) \subset {\cal N}_n \left( S \right) $,
with $ {\cal N}_m \left( S' \right) \not= {\cal N}_n \left( S \right)
$, such that $ g: {\cal N}_m \left( S' \right) \longrightarrow \IN $
is non-bounded.

\noindent {\it Proof}: The neighborhood $ {\cal N}_n \left( S \right)
 $ can be expressed as a finite union of neighborhoods in the
following way: let $ m \geq n + \Delta $, the set
	$$ {\cal W} = \left\{ S' \in {\cal N}_n \left( S \right)
	\vert {\bf S'} \left( - k \right) = {\bf S} \left( - k
	\right) \ \forall k \geq m + \Delta \right\} $$
is finite and
	$$ {\cal N}_n \left( S \right) = \bigcup\limits_{S' \in {\cal
	W}} {\cal N}_m \left( S' \right). $$
Then $ g $ must be non-bounded in at least one of the $ {\cal N}_m
\left( S' \right) $. \hfill \qed

\noindent {\it Proof of theorem 5}: By theorem 2, $ {\cal B^{\infty}}
$ is Hausdorff then by the lemma 4 it remains to prove that for any
covering $ {\cal V} $ $ \exists \ n_0 \in \IN $ $ \ni $ $ \forall S
\in {\cal B^{\infty}} $ $ \exists m \leq n_0 $ $ \ni $ $ {\cal N}_m
\left( S \right) \subset U_i $ for some $ U_i \in {\cal V} $. By
contradiction let us suppose that there does not exist such $ n $.
Then there exists a covering $ {\cal V} $ such that $ \forall n \in
\IN $ $ \exists S \in {\cal B^{\infty}} $ $ \ni $ $ {\cal N}_m \left(
S \right) \subset U_i $ for some $ U_i \in {\cal V} $ implies $ m > n
$. This defines a function
	$$ g: {\cal B^{\infty}} \longrightarrow \IN $$
given by $ g \left( S \right) = n $, which is not bounded. Then by
lemma 5, there exists a nested sequence of neighborhoods
	$$ {\cal N}_{n_1} \left( S_1 \right) \supset {\cal N}_{n_2}
	\left( S_2 \right) \supset ... \supset {\cal N}_{n_k} \left(
	S_k \right) \supset ... $$
with $ n_1 < n_2 < ... $, such that $ g $ is non-bounded $ \forall
{\cal N}_{n_k} \left( S_k \right) $. The sequence $ \left\{ S_k
\right\} $ is a Cauchy net, therefore by theorem 1 it converges to an
element of $ {\cal B^{\infty}} $, but that element is not covered by
$ {\cal V} $ since $ g \longrightarrow \infty $, and therefore we
have a con\-tra\-dic- \break tion.  \hfill \qed

	From theorems 3, 4 and 5 we have the following

\noindent {\it Corollary}: $ {\cal B^{\infty}} $ is a Cantor set.

\

\section{3. Chaotic dynamics in $ {\cal B^{\infty}} $.}

	Let us introduce the following notation: by $ S \left( n
\right) \in {\cal B^{\infty}} $ with $ n \in \IZ $ we understand
	$$ S \left( n \right) = \left( {\bf S} \left( n \right),
	{\bf S} \left( n - 1 \right), {\bf S} \left( n - 2
	\right),... \right). \eqno(8) $$
A binary dynamical system in $ {\cal B^{\infty}} $ is a map $F$ that
is induced by {\it non-vanishing} continuous functions $ F_i: {\cal
B^{\infty}} \longrightarrow \IR - \{ 0 \}$, $ i = 1,...,N $, such
that
	$$ S_i \left( n + 1 \right) = \Theta \circ F_i \left( S
	\left( n \right) \right), \eqno(9)) $$
where $ \Theta \left( x \right) = 0, 1 $ for $ x \leq 0 $, $ x > 0 $
respectively. The map $ F: {\cal B^{\infty}} \longrightarrow {\cal
B^{\infty}} $ is given by
	$$ F \left( S \left( n \right) \right) = S \left( n + 1
	\right). \eqno(10) $$
We should stress that due to the fact that $ {\cal B^{\infty}} $ is
totally disconnected it is not difficult to construct continuous
non-vanishing functions $ F_i $ that change sign.

\noindent {\it Property 3}: The dynamical map defined by (10) is
continuous.

\noindent {\it Proof}: $ F_i: {\cal B^{\infty}} \longrightarrow \IR -
\{ 0 \} $ is continuous and $ \Theta: \IR - \{ 0 \} \longrightarrow
\IZ_2 $ is also continuous, so the composition $ \Theta \circ F_i $
is continuous.  \hfill \qed

	Now we will extend the definitions 1 to 5 of Sec.~1 in a
natural way, so they fit into the dynamics generated by $ F: {\cal
B^{\infty}} \longrightarrow {\cal B^{\infty}} $.

\noindent {\it Definition 9}: The map $ F: {\cal B^{\infty}}
\longrightarrow {\cal B^{\infty}} $ has {\it sensitive dependence on
initial conditions} on $ {\cal A} \subset {\cal B^{\infty}} $ if $
\exists n \in \IN $ $ \ni \forall S \in {\cal A} $ and $ \forall
{\cal N}_m \left( S \right) $ $ \exists S' \in {\cal N}_m \left( S
\right) \cap {\cal A}$ and $ k \in \IN $ $ \ni F^k \left( S' \right)
\notin {\cal N}_n \left( F^k \left( S \right) \right) $.

\noindent {\it Definition 10}: $ F: {\cal B^{\infty}} \longrightarrow
{\cal B^{\infty}} $ is {\it topologically transitive} on $ {\cal A
\subset B^{\infty}} $ if for any open sets $ U, V \subset {\cal A} $
$ \exists n \in \IZ $ $ \ni F^n \left( U \right) \cap V \not=
\emptyset $. In the last expression, if $ F $ is noninvertible we
understand the set $ F^{-k}(U) $ as the set of all points
$ S \in {\cal B^{\infty}} $ such that $ F^k(S) \in U $.

\noindent {\it Definition 11}: Let $ {\cal A \subset B^{\infty}} $ be
a compact set. $ F: {\cal A \longrightarrow A} $ is {\it chaotic} on
$ {\cal A} $ if $ F $ has sensitive dependence on initial conditions
and is topologically transitive on $ {\cal A} $.

\noindent {\it Definition 12}: A closed subset $ {\cal M} \subset
{\cal B^{\infty}} $ is called a {\it trapping region} if $ F \left(
{\cal M} \right) \subset {\cal M} $.

\noindent {\it Property 4}: $ F^n \left( {\cal M} \right)$ is compact
and closed $ \forall n \in \IN $.

\noindent {\it Proof}: Since every closed subset of a compact set is
compact, it follows that $ {\cal M} $ is compact and since $ F $ is
continuous $ F^n \left( {\cal M} \right) $ is compact. Since $ {\cal
B^{\infty}} $ is Hausdorff every compact subset of it is closed, so $
F^n \left( {\cal M} \right) $ is closed~${}^{[15]}$.  \hfill \qed

\noindent {\it Definition 13}: The map $ F: {\cal B^{\infty}}
\longrightarrow {\cal B^{\infty}} $ has an {\it attractor} $ \Lambda
\subset {\cal B^{\infty}} $ if there exists a trapping region $ {\cal
M} $ such that
	$$ \Lambda = \bigcap\limits_{n \geq 0} F^n \left( {\cal M}
	\right). $$

\noindent {\it Property 5}: $ \Lambda $ is compact and closed.

\noindent {\it Proof}: $ \Lambda $ is an intersection of closed sets,
so it is closed. Since every closed subset of a compact space $ {\cal
B^{\infty}} $ is compact, it follows that $ \Lambda $ is compact.
\hfill \qed

\noindent {\it Property 6}: The restriction of $ F $ to $ \Lambda $
is completely defined by specifying $ \Lambda $ itself, as follows.
For any point
$ S \in \Lambda $ and $ n \geq 1 $, let $ S_n \in \Lambda $ be such
that $ {\bf S}_n \left( - n - k \right) = {\bf S} \left( - k \right)
$, for $ k \in \IN $. Then $ F \left( S \right) = \left( {\bf S}_n
\left( - n + 1 \right), {\bf S}_n \left( - n \right),... \right) $.

\noindent {\it Proof}: Since $ S \in \Lambda $, for any $ n \in
\IN $ there is an element $ S_n \in \Lambda $ such that $ F^n
\left( S \right) = S_n $. By (9)) $ F $ acts as a down-shift map for
all but the must recent slice, therefore $ {\bf S}_n \left( - n - k
\right) = {\bf S} \left( - k \right) $, for $ k \in \IN $. \hfill
\qed

\noindent {\it Definition 14}: $ \Lambda $ is called a {\it chaotic
attractor} if $ F $ is chaotic on $ \Lambda $.

\noindent {\it Theorem 6}: If $ \Lambda $ is a chaotic attractor then
it is perfect.

\noindent {\it Proof}: By property 5, $ \Lambda $ is closed, it
remains to prove that every point in $ \Lambda $ is an accumulation
point of $ \Lambda $.  By contradiction, let $ S_0 \in \Lambda $ be
an isolated point, then there exists $ n \in \IN $ $ \ni $ $ {\cal
N}_n \left( S_0 \right) \cap \Lambda = \left\{ S_0 \right\} $ then,
by topological transitivity and the fact that $ F^{- 1} $ exits,
$ \Lambda $ consists of an isolated orbit (the orbit of
$ S_0 $) but this violates sensitivity to initial conditions on $
\Lambda $, so $ \Lambda $ could not be a chaotic attractor.  \hfill
\qed

\noindent {\it Theorem 7}: If $ \Lambda $ is a chaotic attractor then
it is a Cantor set.

\noindent {\it Proof}: The theorem follows directly from property 5,
theorem 6 and the fact that a subset of a totally disconnected set is
also totally disconnected. \hfill \qed

\

\section{4. Metric in $ {\cal B^{\infty}} $.}

	The space $ {\cal B^{\infty}} $ is metrizable as we will see
below.  This enforces its utility in using it to model chaotic
dynamical systems.

\noindent {\it Definition 15}: A metric $ d \left( S_1, S_2 \right) $
is called {\it semicausal} if it induces a semicausal topology.

\noindent {\it Definition 16}: A metric over $ {\cal B^{\infty}} $ is
{\it bounded} if $ \exists M \in \IR $ $ \ni \forall S_1, S_2 \in
{\cal B^{\infty}} $ $ d \left( S_1, S_2 \right) < M $.

\noindent {\it Definition 17}: A metric on $ {\cal B^{\infty}} $ is
{\it local in time} if there exist real numbers $ a > 1 $ and $ w >
0 $ such that
	$$ d \left( S_1, S_2 \right) \leq w a^{- n} $$
implies that
	$$ {\bf S}_1 \left( - m \right) = {\bf S}_2 \left( - m \right) $$
$ \forall m \leq n $.

	The following is an example of a semicausal metric in $ {\cal
B^{\infty}} $:
	$$ d \left( S_1, S_2 \right) = \sqrt{\sum_{n = 0}^{\infty} {1
	\over2^n} {d_n \left( S_1 + S_2 \right) \over 1 + d_n \left(
	S_1 + S_2 \right)}} \eqno(11) $$
where
	$$ d_n \left( S \right) \equiv \sum_{i = 1}^N S^i \left( - n
	\right) $$
and the plus sign is defined by (2). Note that $d_n (S_1 + S_2)$ is
the Hamming distance of the binary models ${\bf S}_1(-n)$ and
${\bf S}_2(-n)$. It is easy to
check that (11) satisfies the following properties:

\noindent \item{\it i)} $  d \left( S_1, S_2 \right) \leq \sqrt{{ 2 N
\over 1 + N}} $ $ \forall S_1, S_2 \in {\cal B^{\infty}} $, so it is
bounded.

\noindent \item{\it ii)} $ d $ is semicausal with $ \Delta = 2 $.

\noindent \item{\it iii)} $ d $ is local in time.

	It is possible to define a ``dot'' product in $ {\cal
B^{\infty}} $ given by
	$$ S \cdot S' = \sum_{n = 0}^{\infty} {1 \over2^n} {d_n
	\left( S \times S' \right) \over 1 + d_n \left( S \times S'
	\right)} $$
where the ``cross'' product is defined by (3). Using equations (6)
and (7) the metric (11) can be written as
	$$ d \left( S_1, S_2 \right) = \sqrt{\left( S_1 + S_2 \right)
	\cdot \left( S_1 + S_2 \right)}. $$

\vfill\eject

\section{5. Conclusion.}

	From an initial {\it ansatz}, to replace the usual
idealization of physical states as ``points'' on a differentiable
manifold by another idealization as infinite ``binary histories'', we
proceeded to define a class of topologies which make the truncation
to finite binary histories a valid approximation, in the same sense
that the continuous topology on $\IR^d$ allows one to approximate a
real coordinate by a finite string of digits or bits. With this
topology and the natural Boolean algebra structure, the space of
binary histories was shown to have several interesting properties,
including those of Cantor sets: it is compact, totally disconnected
and yet every point is an accumulation point.

	Continuous dynamical maps on the space of binary histories
can lead to attracting sets within ${\cal B^{\infty}}$, in which case
an attractor is defined in the usual way. The dynamical map is said
to be chaotic on the attractor if it is sensitive to initial
conditions and topologically transitive.

	It is remarkable that the dynamics on the attractor is
uniquely specified by the attractor itself (Property 6): given any
initial state on the attractor,
	$$ S = \{ {\bf S} \left( 0 \right) , {\bf S} \left( - 1
	\right) , \cdots \} \in \Lambda. $$
one finds, for each $ n $, one and only one state $ S_n \in \Lambda
$ such that
	$$ S_n = \{ {\bf S}_n \left( 0 \right) , {\bf S}_n \left( - 1
	\right) , {\bf S}_n \left( - n + 1 \right) , {\bf S} \left( 0
	\right) , {\bf S} \left( - 1 \right) , \cdots \}.$$
Indeed, since $ S \in \Lambda $ and $ F
\left( \Lambda \right) = \Lambda $, one has $ S_n \in \Lambda $; $
S_n $ is unique because $S_n = F(S_{n-1}), S'_n = F(S_{n-1})
\Rightarrow S_n = S'_n$.  The dynamics is then given by $ F(S) = S_1
$, $ F(F(S)) = S_2 $, etc. In other words, to
determine $ F(S) $ it
is sufficient to scan $ \Lambda $ in search of the state $ S_1 $
which is equal to $ S $ downshifted one step in time, with the extra
binary model $ {\bf S}_1 \left( 0 \right) $ on top; in this sense the
dynamical map is related to a shift map, as in symbolic dynamics.

	The relation to symbolic dynamics is suggestive, and one
might wish to regard our formalism as its generalization.  However,
there are some non-trivial differences between chaos on $ {\cal
B^{\infty}} $ and the shift map on the space of itineraries $
\Sigma^N $ of Ref.~[12]. First of all, points of $ {\cal B^{\infty}}
$ represent the history of the binary system towards the past only,
whereas itineraries extend also to $ t \longrightarrow + \infty $.
Secondly, we are describing chaos on a proper subset $ \Lambda $ of a
larger binary space, $ {\cal B^{\infty}} $, so that $ {\cal
B^{\infty}} $ plays the role of ``embedding space''.  In contrast,
the chaotic dynamics on all of $ \Sigma^N $ is homeomorphic to that
which takes place on a Cantor subset $ \Xi $ of the attractor, that
is $ \Xi \subset \Lambda \subset \IR^d $. In other words one needs to
appeal to a homeomorphism to $ \IR^d $ for the embedding space. The
two differences, one related to causality and the other to the
intrinsic nature of the embedding space $ {\cal B^{\infty}} $,
allowed us to develop the formalism of chaos on binary systems
without ever invoking differentiable manifolds, thereby lending
support to our claim that this formalism can be regarded as a
different representation of reality based on binary histories rather
than real variables.

	One could go one step further and suggest that other theories
of physics could be rewritten by thinking of $ {\cal B^{\infty}} $ as
the space of physical states and call ``real'' the elements $ S \in
{\cal B^{\infty}} $ rather than the coordinates on differentiable
manifolds.  Of course that would probably turn out to be rather
inconvenient for most systems; we are only making this outrageous
suggestion to emphasize that the mathematical constructions which
best represent reality are nothing but those which make reality look
simple; in this sense surely ${\cal B^{\infty}}$ is a more
appropriate framework to describe the physical reality of binary
systems than differentiable manifolds!

	The fact that $ {\cal B^{\infty}} $ is a Cantor set is
perhaps not surprising in light of the analogy to the symbolic
dynamics of chaotic maps $f: \IR^d \to \IR^d$. If one regards an $
N-bit $ binary vector $ {\bf S} \left( 0 \right) $ as defining a
partition of phase space into $ 2^N $ disjoint subsets $ {\cal
O}_{\alpha} $, $ \alpha = 1, 2, \cdots, 2^N $, then the specification
of two consecutive binary vectors, $ \left\{ {\bf S} \left( 0 \right)
, {\bf S} \left( - 1 \right) \right\} $ defines a finer partition
into subsets $ {\cal O}_{\alpha} \cap f \left( {\cal O}_{\beta}
\right) $. For example the $ {\cal O}_\alpha $ might be thought of as
a tiling of phase space, in the sense of Berend and Radin~${}^{[16]}$.

Repeating the procedure to finer and finer partitions one
obtains the image of $ {\cal B^{\infty}} $ in real phase space as the
infinite intersection set
$$ \bigcup\limits_{ \left\{ \alpha, \beta, \gamma, \cdots
	\right\} } \ \left( {\cal O}_{\alpha} \cap f\left( {\cal
	O}_{\beta} \right) \cap f^2 \left( {\cal O}_{\gamma} \right)
	\cap \cdots \right), $$
much like the textbook construction of the Cantor set from the
intersections of the intervals $ I = (0, 1) $, $ f \left( I \right) =
\left( 0, {1 \over 3} \right) \cup \left( {2 \over 3}, 1 \right) $,
etc.  Note that one could replace an $ N-bit $ description of the
state for two consecutive time steps by a $ 2N-bit $ description for
a single time step, based on a partition of phase space into $ 2^{2N}
$ disjoint cells $ {\cal O}_{\alpha} \cap f({\cal O}_{\beta}) $; this
indicates that there is an exact renormalization group transformation
relating refinement in space with extension of binary histories
towards the past; this may be an interesting line of investigation to
pursue which might be expected to raise issues of universality along
the lines of Feigenbaum's work~${}^{[17]}$.

	One rather common example of truncated binary histories is
the case of computer models for chaotic time series prediction. In
the so-called ``method of delays''~${}^{[18]}$, the coordinates of a
point in phase space are taken to be the delayed measurements of a
single variable, so that the state vector is given by $ {\bf x} = \{
x(0), x(-1), x(-2), \cdots, x(-T+1) \} $. Here, $ T $ is the
embedding dimension and each coordinate is represented as a $ 128-bit
$ binary word. The Euclidean metric in $ \IR^T $ induces a semicausal
metric on the space of binary histories.

	A priority in the continuation of this work is to further
elucidate the connection between chaos on binary systems and real
chaos. One notes first of all that there cannot be a homeomorphism
between the embedding space $ {\cal B^{\infty}} $ and $ \IR^d $: one
is a Cantor set and the other a differentiable space!  The reason why
real and binary embedding spaces cannot be homeomorphic is that given
any map from the space of binary histories to real phase space, there
is a continuous curve in the latter which takes one across the
boundary which separates binary histories beginning with distinct
binary vectors, $ {\bf S}' \left( 0 \right) \neq {\bf S} \left( 0
\right) $, so the map is discontinuous at the boundary. This fact
should not be regarded as a serious problem, since chaos is defined
on the attractor and there is no impediment to a homeomorphism from
the attractor $ \Lambda \subset {\cal B^{\infty}} $ to a Cantor
subset of $ \IR^d $, or for that matter from $ {\cal B^{\infty}} $ to
a larger Cantor subset of $ \IR^d $. The formal connection between
chaos on the space of binary histories and real chaos is the subject
of ongoing research.

\

\noindent {\bf Acknowledgements}

\noindent The authors would like to thank Chris Stephens
for helpful comments and discussions.

\vfill\eject

\centerline{{\ftit References}}

\noindent \item{[1]} Martinelli, F. and Olivieri, E.: ``Approach
to equilibrium of Glauber dynamics in the phase region II: The
general case''. Commun. Math. Phys. {\bf 161} No. 3, 487-514 (1994)

\noindent \item{[2]} Comets, F. and Neveu, J. ``The Sherrington-
Kirkpatrick model of spin glasses and stochastic calculus: The
high-temperature case''. Commun. Math. Phys. {\bf 166} No. 3,
549-564 (1995)

\noindent \item{[3]} Glimm, J. and Jaffe, A. ``Quantum physics:
A functional point of view''. New York: Springer 1981

\noindent \item{[4]} Cecotti, S. and Vafa, C. ``Ising model and
$N = 2$ supersymmetric theories''. Commun. Math. Phys. {\bf 157} No. 1,
139-178 (1993)

\noindent \item{[5]} Neves, E.J. and Schonmann, R.H.: ``Critical
dropletsand metastability for a Glauber dynamics at very low
temperature''. Commun. Math. Phys. {\bf 137}, 209-230 (1991)

\noindent \item{} Schonmann, R.H.: ``The pattern of escape from
metastability of a stochastic Ising model''. Commun. Math. Phys.
{\bf 147}, 231-240 (1992)

\noindent \item{[6]} Hertz J., Krogh A. and Palmer R.G. ``
Introduction to the Theory of Neural Computation''.
Redwood City, CA: Addison-Wesley, 1991

\noindent \item{} Zertuche, F., L\'opez-Pe\~na, R. and Waelbroeck,
H. ``Recognition of temporal sequences of patterns with state-
dependent synapses''. J. Phys. A {\bf 27}, 5879-5887 (1994)

\noindent \item{[7]} Bobenko, A., Bordemann, M., Gunn, C. and
Pinkall, U.: ``On two integrable cellular automata''. Commun.
Math. Phys. {\bf 158}, 127-134 (1993)

\noindent \item{[8]} Wolfram, S.: ``Statistical Mechanics of
Cellular Automata'', Rev. Mod. Phys. 55 No. 3, 601-644 (1983)

\noindent \item{[9]} G. Weisbuch. ``Complex Systems Dynamics''.
Addison Wesley, Redwood City, CA. (1991).

\noindent \item{[10]} Chaitin, G.J. in: ``Guanajuato Lectures on
Complex Systems and Binary Networks''. To be published in the
Springer Verlag Lecture Notes series. Eds. R. L\'opez Pe\~na, R.
Capovilla, R. Garc\'\i a-Pelayo, H. Waelbroeck and F. Zertuche.
(1995).

\noindent \item{} Chaitin, G.J.: ``Randomness and Complexity in Pure
Mathematics''. Int. J. Bifurc. Theory and Chaos {\bf 4}, 3-15 (1994)

\noindent \item{[11]} Guckenheimer, J. and Holmes, P.: ``Nonlinear
oscillations, dynamical systems, and bifurcations of vector fields''.
New York: Springer Verlag 1983

\noindent \item{[12]} Wiggins, S.: ``Dynamical Systems and Chaos''.
New York: Springer-Verlag 1990

\noindent \item{[13]} Choquet-Bruhat, Y. and DeWitt-Morette, C.: ``
Analysis, Manifolds and Phys\-ics''. Netherlands: North-Holland 1977

\noindent \item{[14]} ``Encyclopedic Dictionary of Mathematics''
(second edition), by the Mathematical Society of Japan, edited by
Kiyoshi It\^o. Cambridge Massachussetts: MIT Press 1987

\noindent \item{[15]} Munkres, J.R.: ``Topology a First Course''.
New Jersey: Prentice-Hall 1975

\noindent \item{[16]} Berend, D. and Radin, C.: `` Are There Chaotic
Tilings ''. Commun. Math. Phys. {\bf 152} No. 2, 215-219 (1993)

\noindent \item{[17]} Feigenbaum, M.J., Kadanoff, L.P. and Shenker,
S.J.: ``Quasiperiodicity in Dissipative Systems: a Renormalization
Group Analysis''. Physica {\bf 5D}, 370-386 (1982)

\noindent \item{[18]} Tsonis, A.A.: ``Chaos: From Theory to
Applications''. New York: Plenum Press 1992

\vfill \eject

\centerline{{\ftit Figure Captions}}

\noindent \item{[1]} A state in the space of binary histories,
$ S \in {\cal B^{\infty}} $, is a succession of $N - bit$ binary
words giving an approximate description, or ``model'' of the system
at times $ t = 0, -1, -2, \cdots $.

\noindent \item{[2]} A state $S'$ in the neighborhood ${\cal
N}_{n}^{\Delta}(S)$ has the same binary words as $S$ for the slices
$t = 0, -1, \cdots, -n$ and can have any binary word at all beyond $t
= -n - \Delta$. In between these two bounds, the differences which
are allowed for $ S' \in {\cal N}_{n}^{\Delta}(S) $ characterize the
particular semicausal topology.

\bye